\documentclass[preprint,prl,aps,showpacs]{revtex4}
\begin{document}
%\def\sqr#1#2{{\vcenter{\hrule height.#2pt\hbox{\vrule width.#2pt
%height#1pt \kern#1pt \vrule width.#2pt}\hrule height.#2pt}}}
%\def\square{\mathchoice\sqr64\sqr64\sqr{4.2}3\sqr{3.0}3}
%\draft
\newcommand{\nd}{\noindent}
\newcommand{\be}{\begin{equation}}
\newcommand{\ee}{\end{equation}}
\newcommand{\ben}{\begin{eqnarray}}
\newcommand{\een}{\end{eqnarray}}
\newcommand{\nn}{\nonumber \\}
\newcommand{\ii}{\'{\i}}
\newcommand{\pp}{\prime}
\newcommand{\expq}{e_q}
\newcommand{\lnq}{\ln_q}
\newcommand{\quno}{q-1}
\newcommand{\qunoinv}{\frac{1}{q-1}}
\newcommand{\tr}{{\mathrm{Tr}}}

\draft

%@twocolumnfalse\endcsname
\title{Fisher information, Wehrl entropy,  and Landau Diamagnetism}

\author{$^1$S.~Curilef}

\author{$^2$F.~Pennini}
%\thanks{E-mail:~pennini@fisica.unlp.edu.ar}

\author{$^2$A.~Plastino}
%\thanks{E-mail:~plastino@fisica.unlp.edu.ar}
\address{$^1$Departamento de F\'{\i}sica, Universidad Cat\'olica del Norte, \\
 Av. Angamos 0610, Antofagasta, Chile         \\ $^2$Instituto de F\'{\i}sica La Plata (IFLP)\\
 Universidad Nacional de La Plata (UNLP) and Argentine National
 Research Council (CONICET)\\ C.C.~727, 1900 La Plata, Argentina}

%\date{\today}

\begin{abstract}
Using information theoretic quantities like the Wehrl entropy and
Fisher's information measure we study the thermodynamics of the
problem leading to Landau's diamagnetism, namely,  a free spinless
electron in a uniform magnetic field. It is shown that such a
problem can be ``translated" into that of the thermal harmonic
oscillator. We discover a new Fisher-uncertainty relation, derived via the Cramer-Rao inequality,
 that involves phase space localization
 and energy fluctuations.

%%03.65.Sq  Semiclassical theories and  aplications
%%75.20.-g  Diamagnetism, paramagnetism, superparamagnetism
%%05.70.-a  Thermodynamics
%%03.67.-a  quantum information

 \end{abstract}
 \maketitle
 \newpage

 %%%%%%%%%%%%%%%%%%%%%%%%%%%%%%%%%%%%%%%%%%%%%%%%%%%%%%%%%%%%%%%
\section{Introduction}
 %%%%%%%%%%%%%%%%%%%%%%%%%%%%%%%%%%%%%%%%%%%%%%%%%%%%%%%%%%%%%%%
\nd The last years have witnessed a great deal of activity revolving
around physical applications of Fisher's information measure (FIM) $I$
(as a rather small sample, see for  instance,
  \cite{Frieden,roybook, Renyi,FPPS,Incerteza}).  Frieden and Soffer
  \cite{Frieden} have shown that Fisher's information measure provides one
 with a  powerful variational principle,
  the extreme physical information  one,  that yields most of the canonical
 Lagrangians
   of theoretical physics  \cite{Frieden,roybook}.
  Additionally, $I$ has been shown to provide an interesting characterization
  of
  the ``arrow of time'', alternative to the one associated with Boltzmann's
 entropy
  \cite{pla2,pla4}.

  For our present purposes, the  point to emphasize is that equilibrium thermodynamics can
  be entirely based upon Fisher's
measure (via a kind of ``Fisher-MaxEnt"), that exhibits definite
advantages over conventional text-book  treatments \cite{nuestro}.
Evaluating $I$ for a given system is tantamount to possessing {\it complete}
thermodynamic information about it \cite{nuestro}.

Unravelling the multiple FIM facets and their
 links to physics should be of general interest to a vast audience.
 Our subject here is the thermodynamics of Landau's diamagnetism.
 We show, using   FIM that, at 
 temperature $T$, the pertinent physics reduces to that of a
 thermal harmonic oscillator whose frequency is the cyclotron one of the magnetic problem. In doing so,
 a new Fisher-uncertainty relation involving phase space localization
 and energy fluctuations is discovered.

%%%%%%%%%%%%%%%%%%%%%%%%%%%%%%%%%%%%%%%%%%%%%%%%%%%%%%%%
\subsection{Wehrl entropy and Husimi distribution}
%%%%%%%%%%%%%%%%%%%%%%%%%%%%%%%%%%%%%%%%%%%%%%%%%%%%%%%%%
 \label{back}
Quantum-mechanical
  phase-space distributions expressed in terms of the celebrated
  coherent states $\vert z \rangle$ of the harmonic oscillator,
   have been proved to be useful in different contexts ~\cite{Glauber,klauder,Schnack} .
     Particular reference is to be made to  the illuminating work of Andersen and
      Halliwell \cite{PRD2753_93}, who discuss, among other things, the concepts of Husimi distributions and Wehrl entropy.
 Coherent states are   eigenstates of a general
annihilation operator $a$, appropriate for the problem at hand, i.e.,
  $ a\vert z \rangle=z\vert z \rangle$
~\cite{Glauber,klauder,Schnack}. In the special case of the harmonic
  oscillator, for instance, one has
   \ben  \label{z} \mathcal{H}_o &=& \hbar
    \omega \,[ a^{\dagger}  a + 1/2]\,\,
    = i({2\hbar \omega
  m})^{-1/2} p + (m\omega/2\hbar)^{1/2} x \cr
  z&=&(m\omega/2\hbar)^{1/2}x + i({2\hbar \omega
  m})^{-1/2}p.
\een

     Coherent states are often employed together with the concept of Wehrl
  entropy $W$~\cite{wehrl,PRD2753_93,lieb}, a special instance
  of Shannon's logarithmic information measure that constitutes a powerful tool in statistical physics.
  $W$ is defined as
  \be
  W=-\int \frac{\mathrm{d}x\, \mathrm{d}p}{2 \pi \hbar} \mu(x,p)\, \ln
  \mu(x,p),\label{i1}\ee
    where $\mu(x,p)=\langle z| \rho|z\rangle$
  is the ``semi-classical'' phase-space distribution function associated to the
  density matrix $\rho$ of the system~\cite{Glauber,klauder,Schnack}. The distribution  $\mu(x,p)$ is
  normalized in the fashion $ \int (\mathrm{d}x\, \mathrm{d}p/2 \pi
 \hbar)\,
  \mu(x,p)=1,
  $
  and is often referred to as the Husimi
  distribution~\cite{husimi}. The distribution $\mu(x,p)$ is a Wigner
  function smeared over an $\hbar$ sized region of phase space \cite{PRD2753_93}.
  The smearing renders $\mu(x,p)$ a positive function, even is the Wigner distribution does not have
   such a character. The semi-classical Husimi probability distribution refers to a special type of probability:
    that for simultaneous but approximate location of position and momentum in phase space~\cite{PRD2753_93}.

  The usual treatment of equilibrium in statistical mechanics makes use of
the celebrated  Gibbs' canonical distribution,  whose
  associated, ``thermal''
   density matrix is given by  \newline $ \rho=Z^{-1}e^{-\beta  \mathcal{H}}$,
  with $Z=Tr(e^{-\beta \mathcal{H}})$  the partition function, $\beta=1/kT$
   the inverse  temperature ($T$),
  and $k$ the Boltzmann constant. Our present Husimi functions
  will be constructed with such a $\rho$.
  In order to conveniently write down an expression for
  $W$ one considers, for the pertinent   Hamiltonian $\mathcal{H}$,
  its eigenstates   $|n\rangle$ and eigen-energies $E_n$, because one can always write~\cite{PRD2753_93}
\be \mu(x,p)=\langle z| \rho|z \rangle=\frac{1}{Z} \sum_{n}
e^{-\beta
 E_n}|\langle z|n\rangle|^2.\label{husimi}
\ee A useful route to $W$ starts then with (\ref{husimi}) and
continues with (\ref{i1}).
%%%%%%%%%%%%%%%%%%%%%%%%%%%%%%%%%%%%%%%%%%%%%%%%%%%%%%%%%%%%%%%%%%%%%%%%%%
\subsection{Electron without spin  in a uniform magnetic field}
%%%%%%%%%%%%%%%%%%%%%%%%%%%%%%%%%%%%%%%%%%%%%%%%%%%%%%%%%%%%%%%%%%%%%%%%%%
 Consider the kinetic momentum
\be
\overrightarrow{\pi}=\overrightarrow{p}+ \frac{e}{c} \overrightarrow{A},\ee
of a particle of charge $e$,  mass $m_r$, and  linear momentum $\overrightarrow{p}$,
subject to the action of a vector potential
 $\overrightarrow{A}$. These are the essential ingredients of
  the   well-known Landau model for diamagnetism: a spinless electron in a
magnetic field of intensity $H$ (we follow the presentation of
Feldman~{et al.}
 \cite{Feldman}).  The Hamiltonian is  \cite{Feldman} \be
\mathcal{H}=\frac{\overrightarrow{\pi}\cdot
\overrightarrow{\pi}}{2m_r}, \ee  and
the magnetic field is
$\overrightarrow{H}=\overrightarrow{\nabla}\times\overrightarrow{A}$.
The vector potential is chosen in the symmetric gauge as $\overrightarrow{A}=(-H y/2,H
x/2,0)$, which corresponds to a uniform magnetic field along the
$z-$direction. One also  needs the step operators  \cite{Feldman} \be
\pi_{\pm} =p_x \pm i p_y \pm \frac{i \hbar}{2l^2}(x \pm i y). \ee
Motion along the $z-$axis is free \cite{Feldman}.
For the transverse motion, the Hamiltonian specializes to
\cite{Feldman} \be \mathcal{H}_t=\frac{\pi_+\pi_-}{2m_r}+\frac12\,
\hbar \Omega. \ee Two important quantities characterize the problem,
namely, $\Omega=eH/m_r c$, the cyclotron frequency and the length
$l=(\hbar c/e H)^{1/2}$ \cite{lippmann}. The pertinent eigenstates
$|N,m\rangle$ are determined by two quantum numbers: $N$ (associated to the energy) and $m$ (to
the $z-$ projection of the angular momentum). As a consequence, they
 are simultaneously eigenstates of both $\mathcal{H}_t$ and the
angular momentum operator $L_z$  \cite{Feldman}, so that \be
\mathcal{H}_t|N,m\rangle=\left(N+\frac12\right)\hbar \Omega\,
|N,m\rangle = E_N |N,m\rangle \label{transv}\ee and \be
L_z|N,m\rangle=m\hbar|N,m\rangle. \label{transvz} \ee Notice that
the eigenvalues of $L_z$ are not bounded by below ($m$ takes the values $-\infty, \ldots, -1,0,1,\ldots,N$)
 \cite{Feldman}.
This agrees with the fact that  the energies $(N+1/2)\hbar \Omega$
are infinitely degenerate \cite{lippmann}. Moreover, $L_z$ is not an
independent constant of the motion~\cite{lippmann}.

We face a bi-dimensional phase-space problem. The corresponding four
phase-space variables can conveniently be called $x$, $y$, $p_x$,
and $p_y$, since  $\pi_z$ is a constant of the
motion \cite{lippmann} and the motion along the $z-$axis is that of
a free particle.  The pertinent coherent states
$|\alpha,\xi\rangle$ are defined as the simultaneous eigenstates  of
the two commuting non-Hermitian operators which annihilate the
ground state \cite{Feldman} \ben \pi_- \vert N=0,m=0 \rangle&=&0\cr
X_+ \vert N=0,m=0 \rangle&=&0, \een with \cite{Feldman} \be X_{\pm}=
x-\frac{\pi_y}{m_r \Omega} \pm i \left(y + \frac{\pi_x}{m_r
\Omega}\right), \ee that are called orbit-center coordinate
operators that step only the angular momentum $m$ and not the energy
\cite{Feldman}. We have then
\begin{eqnarray}
\pi_-|\alpha,\xi\rangle&=&\frac{\hbar \alpha}{i l^2}|\alpha,\xi\rangle\label{xi}\\
X_+|\alpha,\xi\rangle&=&\xi |\alpha,\xi\rangle,
\end{eqnarray}
where the above defined quantity $l$ represents  the classical
radius of the ground-state's Landau orbit. Evaluating now $ \langle
\alpha,\xi \vert \pi_+ \pi_- \vert \alpha,\xi \rangle$ we
immediately find
  the modulus squared of
 eigenvalue $\alpha$ as given by \cite{Feldman}  \be
|\alpha|^2=\frac{l^4}{\hbar^2}\,\left\{\left(p_x-\frac{\hbar
y}{2l^2}\right)^2+ \left(p_y+\frac{\hbar
x}{2l^2}\right)^2\right\}.\label{alpha2} \ee
The terms within the brackets (divided by $2m_r$) yield the classical energy $\mathcal{E}_{mag}$
of an electron in a uniform magnetic field.
As noted in ~\cite{Feldman}, the  modulus of both $\alpha$ and $\xi$
has dimensions of length.

After expanding the states $|\alpha,\xi\rangle$ in the complete set
of energy eigenfunctions $|N,m\rangle$ given above, and conveniently using Eqs. (3.4) and (3.6) of
\cite{Feldman}, we immediately obtain \be
|\langle
N,m|\alpha,\xi\rangle|^2=\frac{|\alpha|^{2N}|\xi|^{2(N-m)}}{(2l^2)^N
N!\,
 (2l^2)^{N-m} (N-m)!}\,e^{-(|\alpha|^2+|\xi|^2)/2l^2}.\label{Nm}
\ee

Our coherent states $|\alpha,\xi\rangle$ satisfy the closure
relation \cite{Feldman} \be \,\int \frac{\mathrm{d}^2\alpha\,
\mathrm{d}^2\xi}{4\pi^2l^4} \,|\alpha,\xi\rangle\langle
\alpha,\xi|=1, \ee as expected.
%%%%%%%%%%%%%%%%%%%%%%%%%%%%%%%%%%%%%%%%%%%%%%%
\section{Husimi distribution}
%%%%%%%%%%%%%%%%%%%%%%%%%%%%%%%%%%%%%%%%%%%%%%%%

We begin at this point {\it our present endeavor},
i.e., introducing thermodynamics into the model of the preceding Section,
 by calculating the appropriate Husimi
distribution (\ref{husimi}) that our model requires. Such
distribution adopts the appearance \be
\mu(x,p_x;y,p_y)=\frac{1}{Z} \sum_{N=0}^{\infty}
\sum_{m=-\infty}^N \, e^{-\beta
 E_N}|\langle N,m|\alpha,\xi\rangle|^2. \label{husimi2}
\ee

Using  (\ref{Nm}) one can rewrite the above expression in the
fashion \be \mu(x,p_x;y,p_y)=\frac{e^{-\beta \hbar
\Omega/2}}{Z}\,e^{-(|\alpha|^2+|\xi|^2)/2l^2}\,
\sum_{N=0}^{\infty}\,\frac{|\alpha|^{2N}|\xi|^{2N}e^{-\beta \hbar
\Omega N}}{(2l^2)^{2N} N!}\, \sum_{m=-\infty}^{N}\,\left(\frac{2
l^2}{|\xi|^2}\right)^m\frac{1}{(N-m)!},\label{mu2} \ee and pass to
the  evaluation of the sum \be \sum_{m=-\infty}^{N}\,\left(\frac{2
l^2}{|\xi|^2}\right)^m\frac{1}{(N-m)!}= \left(\frac{|\xi|^2}{2
l^2}\right)^{-N}\,e^{|\xi|^2/2l^2}. \ee

This  last result is now replaced  into (\ref{mu2}) so as to arrive
at \be \mu(x,p_x;y,p_y)=\frac{e^{-\beta \hbar
\Omega/2}}{Z}\,e^{-|\alpha|^2/2l^2}\,
\sum_{N=0}^{\infty}\,\left[\frac{|\alpha|^2}{2l^2}\,e^{-\beta \hbar
\Omega} \right]^N\, \frac{1}{N!}, \ee which immediately leads to the
desired Husimi result we were looking for (our first new result),
namely, \be \mu(x,p_x;y,p_y)=\frac{e^{-\beta \hbar
\Omega/2}}{Z}\,e^{-(1-e^{-\beta \hbar \Omega})|\alpha|^2/2l^2}
.\label{mu3} \ee

Feldman~{\it et al.} have given the pertinent partition function $Z$ that we need here, for
  a particle in a
cylindrical geometry (length $L$ and radius $R$), oriented along the magnetic field.
One has $Z_{perp}Z_{parall}$, where $Z_{parall}$ is the usual partition function for one-dimensional free
 motion $Z_{parall}=(L/\hbar)(2\pi m_r k T)^{1/2}$~\cite{Feldman}.
 $Z$ has  the form~\cite{Feldman} \be Z=V \frac{(2 \pi
m_r k T)^{1/2}}{h}\frac{m_r \Omega}{4 \pi \hbar}\frac{1}{\sinh(\beta
\hbar \Omega/2)}. \ee Using it  we can easily recast $\mu(x,p_x;y,p_y)$ as \be
\mu(x,p_x;y,p_y)=\frac{4\pi^2\hbar^2}{V m_r \Omega (2\pi m_r k
T)^{1/2}} \,(1-e^{-\beta \hbar \Omega})\,e^{-(1-e^{-\beta \hbar
\Omega})|\alpha|^2/2l^2} .\label{mu33} \ee This last expression is
not yet normalized (the pertinent normalization integral equals $2\pi
\hbar/(L\sqrt{2 \pi m_r k T})$, with $L$ the length of the sample).
This can be remedied by  scaling the above Husimi distribution.
We proceed in two steps. First we define \be \varphi(x,p_x;y,p_y)=\frac{V m_r \Omega
(2\pi m_r k T)^{1/2}}{4\pi^2\hbar^2}\,\mu(x,p_x;y,p_y) \ee and write
\be \varphi(x,p_x;y,p_y)=(1-e^{-\beta \hbar
\Omega})\,e^{-(1-e^{-\beta \hbar \Omega})|\alpha|^2/2l^2}.
\label{phi3} \ee Although this  is not yet normalized, it is dimensionless.
Now the corresponding normalization integral yields  $A m_r \Omega/(2 \pi \hbar)$.
Finally, the normalized distribution is, of course, \be \phi(x,p_x;y,p_y) =
\frac{2 \pi \hbar}{A m_r \Omega} \left(1-e^{-\beta \hbar
\Omega}\right)\,e^{-(1-e^{-\beta \hbar \Omega})|\alpha|^2/2l^2}.
\label{phi4} \ee
 Obviously, we
write now the Wehrl entropy in terms of the distribution function
$\phi(x,p_x;y,p_y)$ and get \be W=-\int \frac{\mathrm{d}^2\alpha
\mathrm{d}^2\xi}{4\pi^2l^4}\,\phi(x,p_x;y,p_y)\,\ln
\phi(x,p_x;y,p_y),\ee so that, after
 replacing  (\ref{phi4}) into $W$ we find \be
W=1-\ln (1-e^{-\beta\hbar\Omega})- \ln{\left(\frac{2\pi l^2}{A}\right)},\label{Wehrl} \ee where we have
used the following result given in ~\cite{Feldman} \be \int
\frac{\mathrm{d}^2\alpha
\mathrm{d}^2\xi}{4\pi^2l^4}\,e^{-(1-e^{-\beta \hbar
\Omega})|\alpha|^2/2l^2}=\frac{A \mu \Omega}{2 \pi \hbar}\,
\frac{1}{1-e^{-\beta \hbar \Omega}}.\label{integra} \ee  $W$  depends
 on the sample's dimensions via the third term in (\ref{Wehrl}).  The  effect of the
magnetic field is reflected via $\Omega$. The important point is the
following: the present Wehrl measure is, save for the above mentioned (constant) third term,
 {\it identical} to that of
an harmonic oscillator of frequency $\Omega$ at the temperature $T$
\cite{Pennini1}. This constitutes our second original (present) contribution.
 It is to be pointed out that this result
confirms an hypothesis made 10 years ago in \cite{PRD2753_93}, whose authors conjectured that the form
(\ref{Wehrl}) found for the harmonic oscillator could be of a rather general character.

\section{Fisher's information measure}

  \noindent   R.~A.
  Fisher advanced, already in the twenties, a quite interesting information measure
  (for a detailed study see~\cite{Frieden,roybook}).
  Consider a $\theta\,-\,{\bf z}$ ``scenario" in which we deal
  with a system specified by a physical
  parameter $\theta$,  while ${\bf z}$ is a stochastic variable $({\bf z}\,\in\,\Re^{M})$
  and
  $f_\theta({\bf z})$ the probability density for ${\bf z}$
  (that depends also on  $\theta$).  One makes a
  measurement of
   ${\bf z}$ and
  has to best infer $\theta$ from this  measurement,
   calling the
    resulting estimate $\tilde \theta=\tilde \theta({\bf z})$.
The question is how well $\theta$ can be determined. Estimation
theory~\cite{roybook}
   states that the best possible estimator $\tilde
   \theta({\bf z})$, after a very large number of ${\bf z}$-samples
  is examined, suffers a mean-square error $\varepsilon^2$ from $\theta$ that
  obeys a relationship involving Fisher's $I$, namely, $I$$\varepsilon^2=1$,
  where the Fisher information measure $I$ is of the form
  \be
  I(\theta)=\int \,\mathrm{d}{\bf z}\,f_\theta({\bf z})\left\{\frac{\partial \ln f_\theta({\bf z})}{
  \partial \theta}\right\}^2  \label{ifisher}.
  \ee
  \noindent  This ``best'' estimator is the so-called {\it efficient} estimator.
  Any other estimator exhibits a larger mean-square error. The only
  caveat to the above result is that all estimators be unbiased,
  i.e., satisfy $ \langle \tilde \theta({\bf z}) \rangle=\,\theta
  \label{unbias}$.   Fisher's information measure has a lower bound:
 no matter what parameter of the system  one chooses to
  measure, $I$ has to be larger or equal than the inverse of the
  mean-square error associated with  the concomitant   experiment.
  This result,
  \be I\,\varepsilon^2\,\ge \,1, \label{rao}\ee is referred to as the
  Cramer--Rao bound \cite{roybook}. The uncertainty principle can be regarded as a special instance of
  (\ref{rao}) \cite{roybook}. One often speaks of ``generalized" uncertainty relations.

 A particular $I$-case is of great importance:
 that of translation families~\cite{roybook,Renyi},
  i.e., distribution functions (DF) whose {\it
   form} does not change under $\theta$-displacements. These DF
   are shift-invariant (\`a la Mach, no absolute origin for
  $\theta$), and for them
   Fisher's information measure adopts the somewhat simpler appearance
  \cite{roybook}
   \be\label{shift}
  I=\int \,\mathrm{d}{\bf z}\,f({\bf z})\,\left\{\frac{\partial \ln
  f({\bf z})}{
  \partial {\bf z}}\right\}^2.
   \ee
\noindent    Fisher's measure is additive~\cite{roybook}. Here we
deal with the issue of estimating localization in a thermal scenario that
revolves around a four dimensional phase-space, i.e.,  ${\bf z}\equiv
(z_1,z_2,z_3,z_4)$ is a $4$-dimensional vector. Such an
estimation task leads, as shown in \cite{nuestro}, to the
thermodynamics of the problem. Our Fisher measure acquires the appearance   \cite{Pennini1},
  \be \label{aditi} I= \sum_i^4\, I_i=
  \sum_i^4\,\int \,\mathrm{d}{z_i}\,f(z_1,z_2,z_3,z_4)\,\left\{\frac{\partial \ln
  f(z_i)}{
  \partial z_i}\right\}^2. \ee

%%%%%%%%%%%%%%%%%%%%%%%%%%%%%%%%%%%%%%%%%%%%%%%%%%%%%%%%%%%%%%%%
\section{Present application}

Since  $\ln \phi=\ln{(2\pi \hbar/A m_r \Omega)}+ \ln (1-e^{-\beta
\hbar \Omega})-(1-e^{-\beta \hbar \Omega})|\alpha|^2/2l^2$,  the
above result  (\ref{alpha2}) allows for the immediate finding \be
\frac{\partial \ln \phi}{\partial
x}=\frac{1-e^{-\beta\hbar\Omega}}{2\hbar}\left(p_y+\frac{\hbar
x}{2l^2}\right), \ee \be \frac{\partial \ln \phi}{\partial
y}=\frac{1-e^{-\beta\hbar\Omega}}{2\hbar}\left(p_x-\frac{\hbar
y}{2l^2}\right), \ee \be \frac{\partial \ln \phi}{\partial
p_x}=\frac{l^2 (1-e^{-\beta \hbar
\Omega})}{\hbar^2}\left(p_x-\frac{\hbar y}{2l^2}\right), \ee and \be
\frac{\partial \ln \phi}{\partial p_y}=\frac{l^2 (1-e^{-\beta \hbar
\Omega})}{\hbar^2} \left(p_y+\frac{\hbar x}{2l^2}\right). \ee With
the above expressions we can now recast (\ref{alpha2}) in the
fashion \be |\alpha|^2=\frac{2 l^4}{(1-e^{-\beta \hbar
\Omega})^2}\,\mathrm{\cal{A}}, \ee where \be
\mathrm{\cal{A}}=\left(\frac{\partial \ln \phi}{\partial
x}\right)^2+ \left(\frac{\partial \ln \phi}{\partial y}\right)^2+
\frac{\hbar^2}{4l^4}\,\left[\left(\frac{\partial \ln \phi}{\partial
p_x}\right)^2+ \left(\frac{\partial \ln \phi}{\partial
p_y}\right)^2\right]. \ee

We are now in a position to write down the Fisher  measure by
following the prescription (\ref{aditi}) ~\cite{Pennini1,Pennini2}
and then write  \be I=\int \frac{\mathrm{d}^2
\alpha\mathrm{d}^2 \xi}{4 \pi^2 l^4}\,\,\phi(x,p_x;y,p_y)\, l^2
\mathrm{\cal{A}},\label{IT} \ee which, after a little algebra, turns
out to be

\be I=\frac{(1-e^{-\beta \hbar \Omega})^2}{2 l^2}\,
\int \frac{\mathrm{d}^2\alpha \mathrm{d}^2\xi}{4\pi^2l^4}\,
|\alpha|^2\,\phi(x,p_x;y,p_y). \label{flavi}\ee
The integration is performed  by appropriately using the pertinent
derivatives of (\ref{integra}). We finally obtain \be
I=1-e^{-\beta\hbar\Omega}.\label{Fisher} \ee A glance at \cite{Pennini1} tells us that
the above  is just the Fisher measure for the harmonic oscillator, which  constitutes our third original result.
 We can finally compare the information (\ref{Fisher}) with the Wehrl measure
(\ref{Wehrl}), concluding that \be W = 1- \ln I - \ln{\left(\frac{2\pi l^2}{A}\right)}, \ee i.e., they
are complementary informational quantities~\cite{Pennini1}. As a matter of fact, we establish here one of
the few existing {\it direct} Shannon-Fisher links.

For didactic reasons it is now convenient to focus attention on the quantity
 $\vert \alpha \vert^2= 2m_r (l^4/\hbar^2) \mathcal{E}_{mag}$, the ``natural variable" of our scenario,
 go back to Eq. (\ref{flavi}), and notice that the integral is just
$\langle \vert \alpha \vert^2\rangle$, i.e., proportional to the semi-classical mean magnetic energy
$\langle \mathcal{E}_{mag} \rangle$ (see the comment that follows Eq. (\ref{alpha2})). In other words,
estimating localization in phase space is for the
present problem equivalent to evaluating the average energy of our electron.
 It is pertinent to ask now about $\vert \alpha \vert$- fluctuations. A quick calculation yields
\be \langle \vert \alpha \vert\rangle^2= \frac{\pi l^2}{2I},\ee and

\be (\Delta\langle \vert \alpha \vert\rangle)^2=  \langle \vert \alpha \vert^2\rangle
- \langle \vert \alpha \vert\rangle^2 = \frac{4-\pi}{2} \,\,\,\frac{l^2}{I}.\ee Out phase space localization
problem becomes intimately linked to  these fluctuations. The ensuing
$(\Delta\langle \vert \alpha \vert\rangle)^2\,I$-product, i.e., the $\vert \alpha \vert$-Cramer-Rao bound (\ref{rao})
(generalized uncertainty principle   \cite{roybook}) is

\be (\Delta\langle \vert \alpha \vert\rangle)^2\,I=\frac{4-\pi}{2}\, l^2= \frac{4-\pi}{2}\,\,\frac{c}{eH}\,\,\hbar, \ee
and we observe: i) as an equal sign is obtained, the estimation is {\it optimal}
in the sense that the lower bound of the inequality (\ref{rao}) is always obtained \cite{roybook}, ii)
the associated uncertainty is {\it independent} of the temperature, and iii) as we increase localization-quality
($I$ increases), the size of $\vert \alpha \vert$-fluctuations, reasonably enough, decreases. A control-parameter, namely,
 the magnetic field intensity $H$, is available. The larger the intensity, the better the overall quality.
 Nature imposes the ultimate control, however, as given by $\hbar$. \newline The difference between $(4-\pi)/2$ and
 $1/2$ (of the order of 0.36) is
 due to the semi-classical character of our treatment.

 We look now for a Cramer-Rao inequality that directly involves the energy $ \mathcal{E}_{mag}$. Things
 will drastically change because to get the energy from $\vert \alpha \vert^2$
 one must divide by $l^4$, which in turn will {\it reverse} the $H-$role.  We immediately find
  \be  \langle \mathcal{E}_{mag} \rangle = \frac{\hbar \Omega}{I},\ee and

\be  \langle \mathcal{E}_{mag}^2 \rangle = 2 \frac{\hbar^2 \Omega^2}{I^2}, \ee
so that for the energy-fluctuation
$\Delta^2\mathcal{E}_{mag}=\langle \mathcal{E}_{mag}^2 \rangle - \langle \mathcal{E}_{mag} \rangle^2 $
one finds
\be  \Delta\mathcal{E}_{mag}\,I= \hbar \Omega= \hbar \frac{eH}{m_rc},  \label{fla1} \ee
which, once again, is independent of $T$. The effect of $H$ is clearly different now, as anticipated.
It is a simple matter to verify that (\ref{fla1}) also gives a localization-energy fluctuations  Cramer-Rao
 uncertainty for the harmonic oscillator. The smaller the energy fluctuations, the better the localization
 estimation via $I$.
\section{Conclusions}
A semi-classical information theory undertaking was tackled here: i) trying to estimate phase-space
location via Fisher information and ii) evaluating the semi-classical Wehrl entropy, for the celebrated
Landau's diamagnetism problem. Evaluating the Fisher measure $I$ appropriate for the problem
yields its thermodynamics \cite{nuestro}. As a summary:
\begin{itemize}
\item Using the coherent states discussed in  Ref. \cite{Feldman} we
have explicitly given the form of the Husimi distribution function
for a spinless electron in a uniform magnetic field (Cf. Eq.
(\ref{mu3})).
\item We have discovered that the Wehrl entropy for Landau's
diamagnetism is, save for a constant term that depends on the size of the sample,
 that of a thermal harmonic oscillator whose
frequency is the cyclotron one.
\item For the corresponding Fisher measure the above similitude becomes identity.
The thermo-statistics of the two problems is thus the same at the
semi-classical level.
\item We confirmed a conjecture made in \cite{PRD2753_93}, in the sense that  the form
(\ref{Wehrl}) could be of a rather general character.
\item An uncertainty relation linking phase space localization with energy fluctuations has been
discovered (Cf. Eq. (\ref{fla1})).
\end{itemize}

{\bf Acknowledgment} One of us (S.C) would like to thank partial
financial support by FONDECYT, grant 1010776.

  \end{document}